\documentclass{PoS}
\usepackage{mathtools}
\usepackage{amssymb,amsmath}

\def\Tr{{\rm Tr}}

\title{Exclusive diffractive processes including saturation effects at next-to-leading order}

\ShortTitle{Exclusive diffractive processes at NLO}

\author{\speaker{Samuel Wallon}\\
        Laboratoire de Physique Th\'eorique (UMR 8627), CNRS, Univ. Paris-Sud, Universit\'e Paris-Saclay, 91405, Orsay, France {\em \&} \\
Sorbonne Universit\'e, Facult\'e de Physique, 4 place Jussieu, 75252 Paris Cedex 05, France\\
        E-mail: \email{samuel.wallon@th.u-psud.fr}}

\author{Renaud Boussarie\\
Physics Department, Brookhaven National Laboratory, Upton, NY 11973, USA\\
Email: \email{rboussarie@bnl.gov}}

\author{Andrey~V.~Grabovsky
\\
Budker Institute of Nuclear Physics, 11, Lavrenteva avenue, 630090, Novosibirsk, Russia {\em \&} \\
Novosibirsk State University, 630090, 2, Pirogova street, Novosibirsk, Russia\\
Email: \email{A.V.Grabovsky@inp.nsk.su}}

\author{Lech Szymanowski
\\
National Centre for Nuclear Research (NCBJ), 02-093 Warsaw, Poland\\
Email: \email{Lech.Szymanowski@ncbj.gov.pl}}

\abstract{In the framework of the QCD shock-wave approach, we review our results on the description
of diffractive production of various final states (jets, meson) at next-to-leading order.
This is applied to exclusive diffractive dijet electroproduction at HERA.}

\FullConference{Light Cone 2019 - QCD on the light cone: from hadrons to heavy ions - LC2019\\
		16-20 September 2019\\
		Ecole Polytechnique, Palaiseau, France}

\begin{document}

The HERA research
program revealed that almost 10\% of the deep
inelastic scattering (DIS) events were shown to contain a
rapidity gap in the detectors between the proton remnants 
and the hadrons coming from the fragmentation region of
the initial virtual photon.
Among these events, exclusive diffractive production of dijets is particularly promising in order to distinguish between a collinear QCD factorized description involving distributions of
partons inside the exchanged Pomeron~\cite{Collins:1997sr}, and a high-energy description in which the Pomeron is directly coupled to the hard subprocess. We here briefly report on this second description, including gluonic saturation within the 
QCD shockwave approach, which we then apply to ZEUS data.

\section{The QCD shockwave approach}

\begin{figure}
\centerline{\includegraphics[width=7cm]{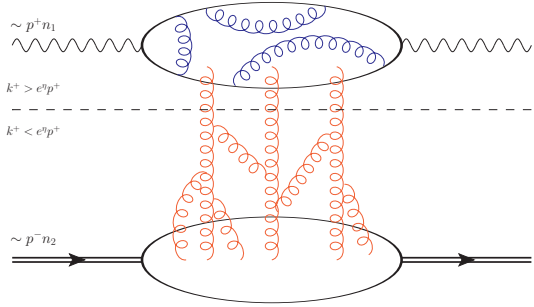}}
\caption{Splitting between quantum and classical parts.}
\label{Fig:rapidity-separation}
\end{figure}

In a balanced frame, e.g. center-of-mass frame (c.m.f), consider a projectile scattering a target respectively flying almost along light-cone directions 
$n_1$ and
$n_2$, with
\begin{equation}
n_1 = \sqrt{\frac{1}{2}}(1,0_\perp,1), \quad n_2 = \sqrt{\frac{1}{2}}(1,0_\perp,-1), \quad (n_1\cdot n_2) = 1\,.
\end{equation}
Introducing lightcone coordinates 
\begin{equation}
x = (x^0,x^1,x^2,x^3) \rightarrow (x^+,x^-,\vec{x}) \, \quad  \text{ with } \quad
x^+ = x_- = (x\cdot n_2) \,,\quad x^- = x_+ = (x \cdot n_1)  
\end{equation}
and a rapidity separation $\eta$ (with $e^{\eta}  \ll 1$), the gluonic field can be split between ``fast'' (quantum part) and ``slow'' (classical part) as illustrated in Fig.~\ref{Fig:rapidity-separation}:
\begin{eqnarray}
\mathcal{A}^{\mu a} (k^+, k^- ,\vec{k}\,) & = & {\color{blue}{A_\eta^{\mu a}\,(|k^+| > e^{\eta}p^+,k^-,\vec{k}\, )}} \qquad \hbox{\color{blue}{quantum part}} \nonumber \\
& + & {\color{red}{b_\eta^{\mu a}(|k^+| < e^{\eta}p^+,k^-,\vec{k}\, )}}\qquad \hbox{\color{red}{\ \ classical part}}. 
\end{eqnarray}
\begin{figure}
\begin{tabular}{ccc}
\includegraphics[width=6.3cm]{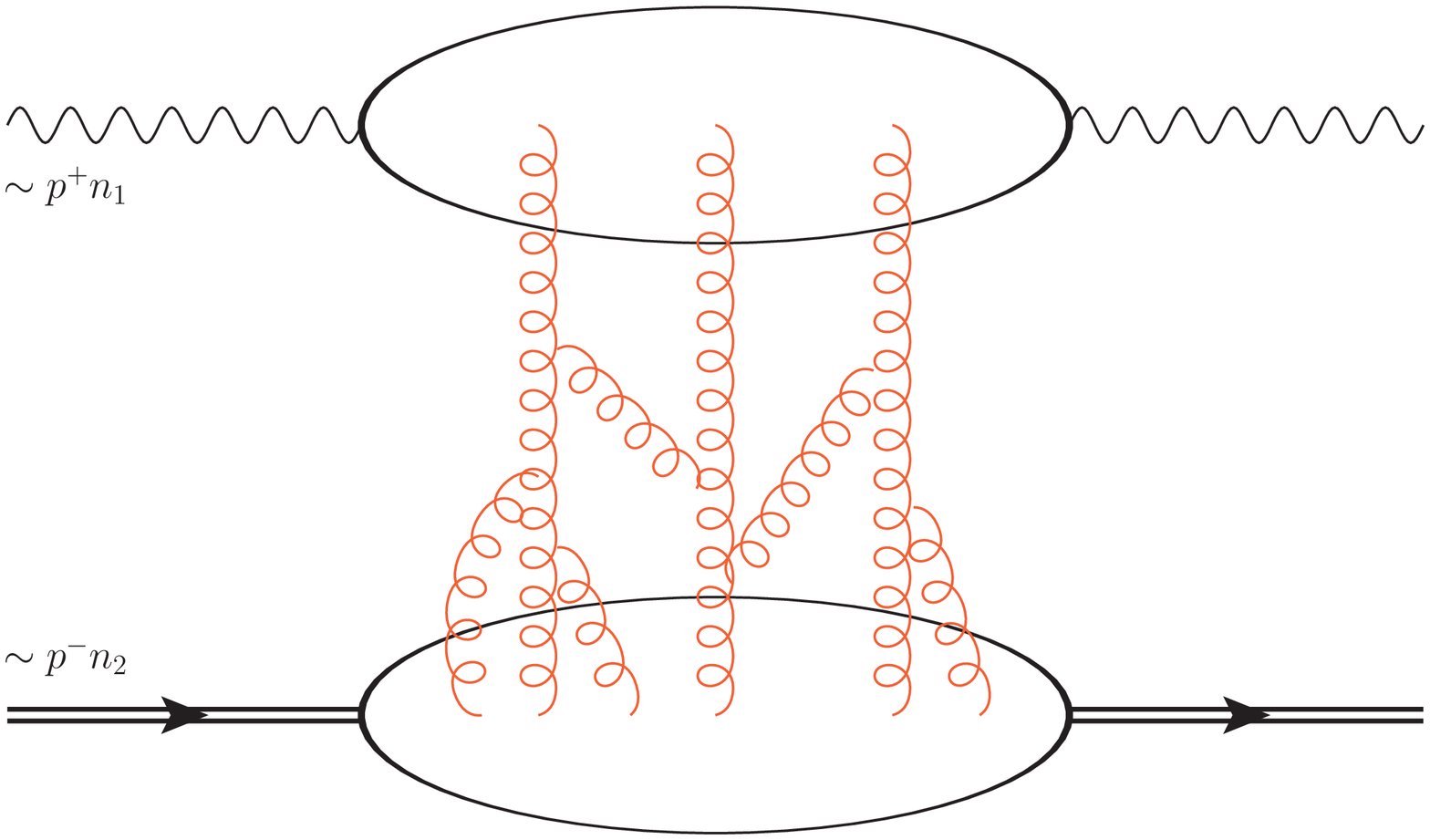}

& \raisebox{2cm}{$\xrightarrow[]{\text{{boost}}}$}
&
\includegraphics[width=6.3cm]{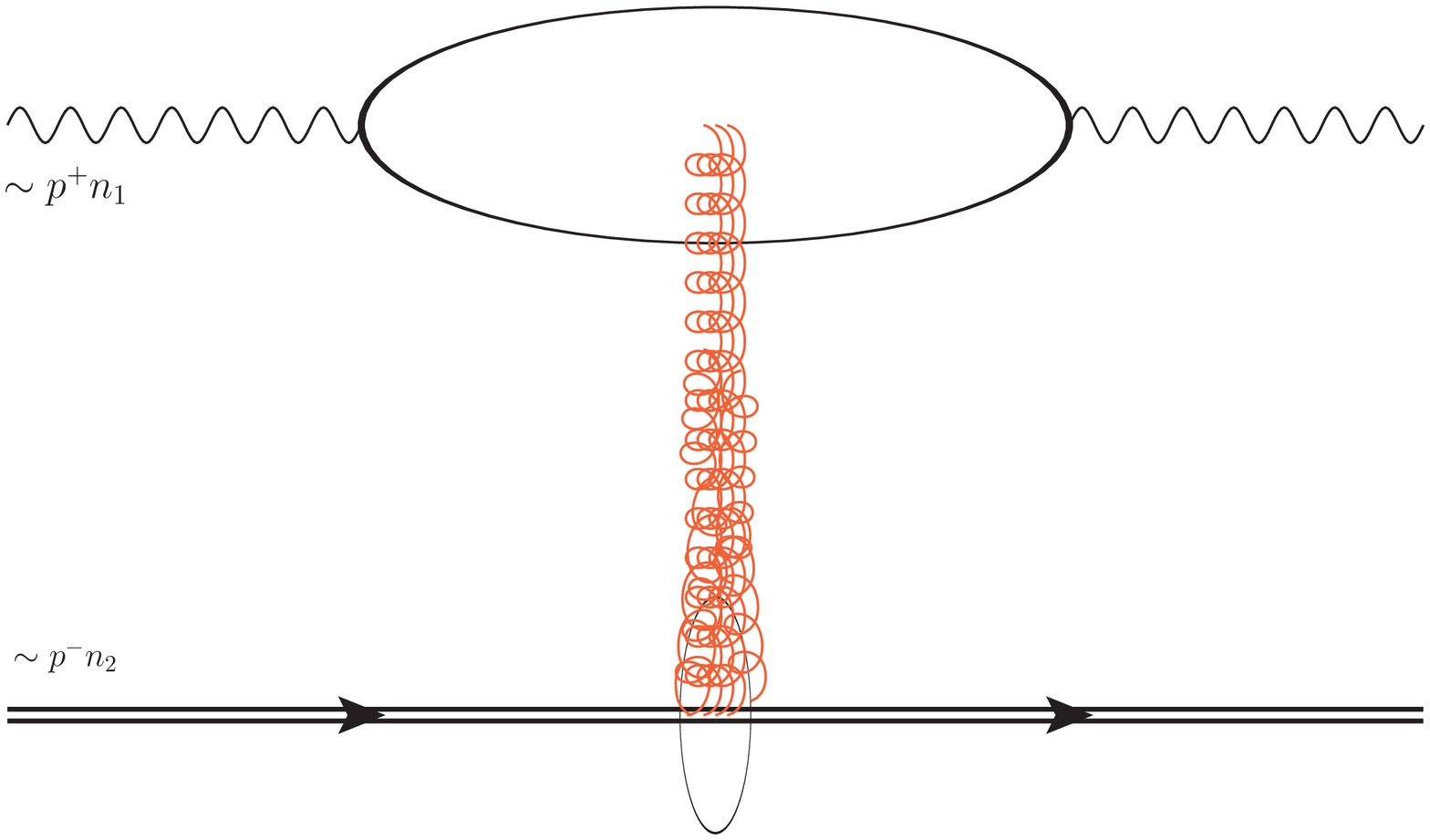}
\\
$b^\mu(x)$ & & $b^-(x)\, n_2^\mu \simeq {\color{red}{\delta(x^+) \, \text{B}(\vec{x}\,) \, n_2^\mu}}$
\end{tabular}
\caption{The shockwave approximation after a large longitudinal boost to the projectile frame.}
\label{Fig:boost}
\end{figure}
In the boosted projectile frame, the classical part $b^\mu$ has a particularly simple Lorentz structure, as illustrated in Fig.~\ref{Fig:boost}.
Multiple interactions with the target can then be resummed into path-ordered
Wilson lines $U_{\bar{z}_i}$ attached to each parton crossing lightcone time $x^+=0:$
\begin{equation}
\label{Uzi}
U^\eta_{\vec{z}_{i}}  =P e^{{ig\int{b_{\eta}^{-}(z_{i}^{+} \! ,\vec{z}_{i}) } \, dz_{i}^{+}}}\,.
\end{equation}
Finally, a factorized picture arises, which allows for a description of the scattering amplitude as a convolution, in transverse space, of 
matrix elements of the
Wilson line operators acting on
the target states with the impact factor describing the scattering of the projectile off the classical field:
\begin{eqnarray}
\mathcal{A} = \int\!\!  d\vec{z}_1 ... d\vec{z}_n \,\, {\color{blue}{\Phi(\vec{z}_1,...,\vec{z}_n)}} 
{\color{red}{\langle P^\prime | U_{\vec{z}_1}...U_{\vec{z}_n} | P \rangle}} \,. \nonumber
\end{eqnarray}
The
Wilson line operators evolve with $\eta$ through the Balitsky hierarchy~\cite{Balitsky:1995ub}, 
which includes non linear terms responsible for gluonic saturation. Equivalently to this high energy operator expansion, a functional approach has been developped, known as the color glass condensate formulation~\cite{McLerran:1994vd}, governed by the JIMWLK evolution equation~\cite{JalilianMarian:1997jx,JalilianMarian:1997gr,JalilianMarian:1997dw,JalilianMarian:1998cb,Kovner:2000pt,Weigert:2000gi,Iancu:2000hn,Iancu:2001ad,Ferreiro:2001qy}.

Focusing on color-singlet exchange, exclusive diffraction allows one to probe the impact parameter $b_\perp$-dependence of the non-perturbative scattering amplitude. We  went for the first time beyond leading-order (LO) in computing impact factors at next-to-leading order (NLO) in the case of diffractive exclusive dijet~\cite{Boussarie:2014lxa,Boussarie:2016ogo} and light vector meson~\cite{Boussarie:2016bkq} production in arbitrary kinematics. A noticeable outcome is the fact that besides an intermediate color-dipole 
made by the $q\bar{q}$ pair, a configuration made of two dipoles is involved when an additional gluon at NLO goes through the shock-wave.

\section{Exclusive
diffractive dijet electroproduction at HERA}

We investigated the ZEUS diffractive exclusive dijet measurements performed at HERA~\cite{Abramowicz:2015vnu}, see details in Ref.~\cite{Boussarie:2019ero}. 
We denote as $W$ the $\gamma^* P$ total energy in the c.m.f., $Q^2$ the (opposite) $\gamma^*$ virtuality, and $M$ the mass of the diffractive dijet system.
At LO, the $\gamma_L^* P$ cross-section
\begin{equation}
\left.  \frac{d\sigma_{0LL}}{dt}\right\vert _{t=0}=\frac{1}{2(2\pi)^{4}}%
\frac{4\alpha Q_{q}^{2}}{N_{c}}\pi\int dxQ^{2}x^{2}\bar{x}^{2}\int d^{2}%
rK_{0}(\sqrt{x\bar{x}}Qr)^{2}F(\vec{r})^{2}
\end{equation}
is expressed through the forward dipole matrix element 
\begin{equation}
F(z_{\bot})= \left.  
\frac{
\langle P^{\prime}%
(p_{0}^{\prime})|T(\Tr(U^{}_{\frac{z_{\bot}}{2}}U_{-\frac{z_{\bot}}{2}}^{\dag
  })-N_{c})|P(p_{0})\rangle
  }{2\pi\delta(p_{00^{\prime}}^{-})}\right\vert
_{p_{0}\rightarrow p_{0}^{\prime}}
=N_{c}\sigma_{0}(1-e^{-\frac{z^{2}}{4R_{0}^{2}}})\,,
\end{equation}
where we use the Golec-Biernat-W\"usthoff (GBW) parametrization~\cite{GolecBiernat:1998js} in the last equality, therefore including saturation for dipoles of transverse size larger than $R_0$.
Here%
\begin{equation}
\label{R0}
\quad R_{0}=\frac{1}{Q_{0}}\left(  \frac{x_{P}}{a_{0}}\right)  ^{\frac
{\lambda}{2}},
\end{equation}
with
\begin{equation}
\label{def:xP-general}
x_{P}=\frac{Q^{2}+M^2-t}{Q^{2}+W^{2}}\,,
\end{equation}
which describes the fraction of the incident momentum
lost by the proton or carried by the Pomeron exchanged in $t-$channel~\footnote{A detailed analysis using various models including saturation has been recently performed at LO in Ref.~\cite{Salazar:2019ncp}.}

At NLO, besides the LO contribution described by the emission of a $q\bar{q}$ pair from an initial virtual photon which goes through the classical gluonic field of the proton,  one should further include configurations in which the dijet system can be made of three partons (real contributions) as well as of two partons with a one loop correction (virtual contributions). The precise way one attributes two and three partons to dijets or trijets configurations goes through a jet algorithm.
ZEUS used the exclusive $k_t-$jet algorithm~\cite{Brown:1991hx}. Let $E_i$, $E_j,$  be the  particle's energies and  $\theta_{ij}$ the relative
angle between them in the c.m.f,
the
distance between two particles is defined as
\begin{equation}
d_{ij}=2\min(E_{i}^{2},E_{j}^{2})\frac{1-\cos\theta_{ij}}{M^2}=\min
\left(\frac{E_{i}}{E_{j}},\frac{E_{j}}{E_{i}}\right)\frac{2p_{i}\cdot p_{j}}{M^2} \,.
\end{equation}
The two particles then belong to one jet if $d_{ij} < y_{cut},$ 
where 
$y_{cut}$ regularizes both soft and collinear singularities. In practice, 
$y_{cut}=0.15$ in ZEUS analysis, and we rely on a small $y_{cut}$ approximation.

The cuts used by ZEUS are 
$
5~{\rm GeV} <Q \,, \ 
5~{\rm GeV} < M_{2jets} < 25~{\rm GeV} \,, \ 
2~{\rm GeV} < p_{\bot\min}\,.
$
At Born level, this removes the aligned jets configurations
$x\lesssim\frac{1}{\max(Q^{2}%
,\,M^2)R_{0}^{2}}\ll1$, 
the leading twist contribution which normally dominates in the GBW saturation model.
Besides, the typical hard scale in
the impact factor is larger than  $p_{\bot\min}^{2} > Q_{s}^{2},$
justifying an expansion in powers of $Q_s$:  
ZEUS experiment is dominated by the linear BFKL regime.
We restrict ourselves to the dominant contributions:

- Born cross section with soft and collinear corrections

- real correction with dipole $\times$ dipole, dipole $\times$ double dipole, and double dipole $\times$ double dipole
contributions for the gluon dipole dijet configuration.

\begin{figure}
\centerline{
\raisebox{0cm}{\includegraphics[angle=0,width=100mm]{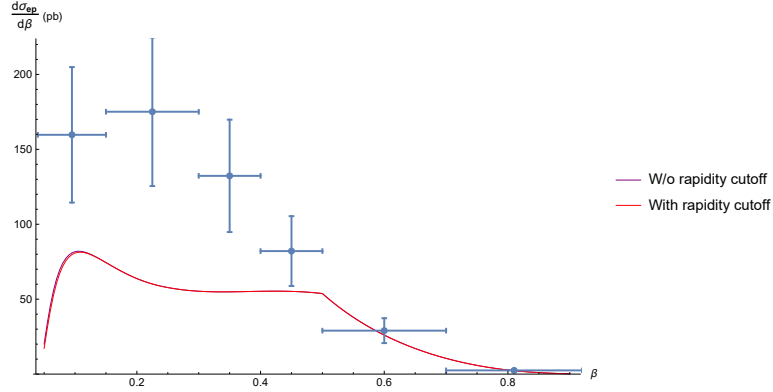}}}%
\caption{Born and total gluon dipole contributions to cross section.}
\label{Fig:theory-versus-data}
\end{figure}

The sum of these contributions is compared with ZEUS data for cross-section in Fig.~\ref{Fig:theory-versus-data}, as a function 
of the Bjorken variable $\beta$ normalized to the pomeron momentum 
\begin{equation}
\label{def:beta-general}
\beta  =\frac{Q^{2}}{Q^{2}+M^2-t} \simeq \frac{Q^{2}}{Q^{2}+M^2} \quad \text{at small }t.
\end{equation}
One gets a good agreement with data at
 large $\beta$, while at 
 small $\beta$ there is a poor agreement with data, as for the two gluon model of Ref.~\cite{Bartels:1998ea}. We get similar conclusions for the azimuthal distribution of the jets.
 This calls for an inclusion of the remaining nonenhanced contributions (the nonsingular part of the virtual corrections, and the remaining part of the real one).
 
 \section{Conclusion}

 We provided the first full NLO computation of the $\gamma^{(*)} \to {\rm jet} \ {\rm jet}$ and $\gamma^{(*)}_{L,T} \to \rho_L$
impact factors. This can be adapted for twist 3  $\gamma^{(*)}_{L,T} \to \rho_T$ NLO production
in the Wandzura-Wilczek approximation, removing factorization breaking
end-point singularities even at NLO for a process which would not
factorize in a full collinear factorization scheme~\cite{Anikin:2009hk,Anikin:2009bf}. For dijet electroproduction, in the small $y_{cut}$ limit of the exclusive $k_t$-jet algorithm, and for large $\beta$, a good agreement between the
GBW model (in the small $Q_s$ expansion) combined with
our NLO impact factor and ZEUS data is obtained.
This is a good sign that perturbative Regge-like descriptions are
favored with respect to collinear type descriptions.
Finally, one should note that within ZEUS kinematical cuts, the linear BFKL regime dominates, while
EIC should give a direct access to the saturated region.

\section*{Acknowledgments}

This project has received funding from the European Union's Horizon 2020 research and innovation programme under grant
agreement No 824093. The work of R.~B. is supported by the U.S. Department of Energy, Office of Science, Office of Nuclear Physics,
under contract No.DE-SC0012704, and in part by Laboratory Directed Research and Development (LDRD) funds
from Brookhaven Science Associates. The work of L.~S. is supported by the grant 2017/26/M/ST2/01074 of the National Science Center in Poland. He thanks
CNRS, the LABEX P2IO, the GDR QCD. The project is also co-financed by the Polish National Agency for Academic Exchange. The work of A.~V.~G. is supported by the Russian Fund of Basic Research grant 19-02-00690 and the Ministry of Education and Science
of Russia.


\providecommand{\href}[2]{#2}\begingroup\raggedright\endgroup

\end{document}